\documentclass[prd, preprint, nofootinbib]{revtex4}

\usepackage{graphicx}
\usepackage{amsfonts}
\usepackage{latexsym}
\usepackage{amsmath}
\usepackage{amssymb}
\usepackage{epsfig}

\begin{document}

\title{ Gravitational waves from the minimal gauged $U(1)_{B-L}$ model}

\author{Taiki Hasegawa}
 \email{t-hasegawa@particle.sci.hokudai.ac.jp}
 \affiliation{Department of Physics, Hokkaido University, Sapporo 060-0810, Japan}

\author{Nobuchika Okada}
 \email{okadan@ua.edu}
 \affiliation{
Department of Physics and Astronomy, 
University of Alabama, Tuscaloosa, Alabama 35487, USA}

\author{Osamu Seto}
 \email{seto@particle.sci.hokudai.ac.jp}
 \affiliation{Institute for the Advancement of Higher Education, Hokkaido University, Sapporo 060-0815, Japan}
 \affiliation{Department of Physics, Hokkaido University, Sapporo 060-0810, Japan}

%

\begin{abstract}
An additional $U(1)$ gauge interaction is one of the promising extensions of the standard model of particle physics.
Among others, the $U(1)_{B-L}$ gauge symmetry is particularly interesting
 because it addresses the origin of Majorana masses of right-handed neutrinos,
 which naturally leads to tiny light neutrino masses through the seesaw mechanism. 
We show that, based on the minimal $U(1)_{B-L}$ model,
 the symmetry breaking of the extra $U(1)$ gauge symmetry with its minimal Higgs sector in the early universe
 can exhibit the first-order phase transition and hence generate a large enough
 amplitude of stochastic gravitational wave radiation that is detectable in future experiments. 
\end{abstract}


\preprint{EPHOU-19-004} 

\vspace*{3cm}
\maketitle


\section{Introduction}

The standard model (SM) of particle physics based on the gauge group $SU(3)_C\times SU(2)_L \times U(1)_Y$
 successfully describes most of the elementary particle phenomena below the TeV scale. 
Nevertheless, an additional $U(1)$ gauge interaction is one of promising extensions of the SM.
As a minimal extension of the SM, we may consider models based on
 the gauge group $SU(3)_C \times SU(2)_L \times U(1)_Y \times U(1)_{B-L}$~\cite{Mohapatra:1980qe,Mohapatra:1980},
 where the $U(1)_{B-L}$ (baryon number minus lepton number) gauge symmetry is supposed to be broken 
 at a high energy scale.  
In this model with a conventional $U(1)_{B-L}$ charge assignment,
 the gauge and gravitational anomaly cancellations require us to introduce three right-handed (RH) neutrinos
 whose Majorana masses are generated by the spontaneous breakdown of the $U(1)_{B-L}$ gauge symmetry. 
With Yukawa interactions between left-handed and RH neutrinos,
 the observed tiny neutrino masses is naturally explained by the so-called seesaw mechanism  
 with the heavy Majorana RH neutrinos~\cite{SeesawY,SeesawG,SeesawM}. 
If the energy scale of the $U(1)_{B-L}$ symmetry breaking is much higher than the TeV scale as is often assumed,
 it is very difficult for any collider experiments to test the mechanism of the symmetry breaking and
 the RH neutrino mass generation.

In Ref.~\cite{Okada:2018xdh}, two of the present authors showed that
 the $U(1)_{B-L}$ symmetry breaking with a simple extended $U(1)_{B-L}$ Higgs sector at such a high energy
 below about $10^7$ GeV scale can be probed by the detection of a gravitational wave (GW) and
 its energy scale dependence in the spectrum. 
For a similar work, see, e.g., Ref.~\cite{Brdar:2018num}.
Although it has been pointed out that GWs generated by a first-order phase transition 
 at a temperature about $10^7$ GeV could be in reach of future experiments~\cite{Grojean:2006bp,Dev:2016feu,Balazs:2016tbi},
 the possibility has not been explicitly demonstrated
 for the specific high scale $U(1)_{B-L}$ model well motivated by neutrino physics before the work in
 Ref.~\cite{Okada:2018xdh}.\footnote{For studies on GWs generated by a TeV scale $U(1)_{B-L}$ phase transition,
 one may find several papers, for example, Refs.~\cite{Jinno:2016knw,Marzo:2018nov} for the classical conformal
 invariance model~\cite{Iso:2009ss,Iso:2009nw} and Ref.~\cite{Chao:2017ilw} for an extended Higgs potential.
GWs from a second-order $B-L$ phase transition during reheating also have been studied in Ref.~\cite{Buchmuller:2013lra}.}
In this previous work, 
 the Higgs sector was extended so that a trilinear interaction of scalar fields
 in the tree level potential can be introduced as in Ref.~\cite{Chao:2017ilw}, 
 which play a crucial role to trigger the first-order phase transition. 

In this paper, we investigate GWs from the first-order phase transition associated 
 with the spontaneous $U(1)_{B-L}$ gauge symmetry breaking within the minimal model context. 
If a first-order phase transition happens in the early universe,
 the dynamics of bubble collision~\cite{Turner:1990rc,Kosowsky:1991ua,Kosowsky:1992rz,Turner:1992tz,Kosowsky:1992vn}
 followed by turbulence of the plasma~\cite{Kamionkowski:1993fg,Kosowsky:2001xp,Dolgov:2002ra,Gogoberidze:2007an,Caprini:2009yp}
 and sonic waves generate GWs~\cite{Hindmarsh:2013xza,Hindmarsh:2015qta,Hindmarsh:2016lnk}, 
 which can be detected by the future experiments,  
 such as the eLISA~\cite{Seoane:2013qna}, the Big Bang Observer (BBO)~\cite{Harry:2006fi},
 DECi-hertz Interferometer Observatory (DECIGO)~\cite{Seto:2001qf} and Advanced LIGO (aLIGO)~\cite{Harry:2010zz}.

This paper is organized as follows: 
In the next section, we introduce formulas we adopt to describe the spectrum of GWs generated by
 a cosmological first-order phase transition.
In Sec.~\ref{sec:gw}, we describe the minimal $U(1)_{B-L}$ model 
  and then derive the resultant GWs spectrum by estimating the latent heat
 and the transition timescale of the phase transition. 
We also discuss model parameter dependence of the GWs spectrum.
In the last section, we summarize our results.

\section{GW generation by cosmological first-order phase transition}

In this section, we briefly summarize the properties of GWs produced by
 three main GW production processes and mechanisms:
 bubble collisions, turbulence~\cite{Kamionkowski:1993fg}, and
 sound waves after bubble collisions~\cite{Hindmarsh:2013xza}.
See, for instance, Refs.~\cite{Caprini:2018mtu,Mazumdar:2018dfl} for a recent review. 
The GW spectrum generated by a first-order phase transition of a Higgs field
 critically depends on two quantities:
 the ratio of the latent heat energy to the radiation energy density $\rho_\mathrm{rad}$, which is expressed by
 a parameter $\alpha$, and the transition speed $\beta$ defined below.
In this section, we give definitions of those parameters and the fitting formula
 of the GW spectrum.

\subsection{Scalar potential parameters related to the GW spectrum}

A phase transition is induced by a scalar field $\phi$ in the radiation dominated universe with temperature $T$.
At the moment of a first-order phase transition, the latent energy density is given by
\begin{equation}
\epsilon = \left.\left(V - T\frac{\partial V}{\partial T}\right)\right|_{\{\phi_\mathrm{high}, T_{\star}\}} - \left.\left(V -T\frac{\partial V}{\partial T}\right)\right|_{\{\phi_\mathrm{low}, T_{\star}\}}, 
\end{equation}
 where $\phi_{\mathrm{high}(\mathrm{low})}$ denotes the field value of $\phi$ at the high (low) vacuum.  
Here and hereafter, quantities with the subscript $\star$ denote those
 at the time when the phase transition takes place~\cite{Huber:2008hg}. 
On the other hand, the radiation energy density is given by
\begin{equation}
\rho_\mathrm{rad} = \frac{\pi^2 g_*}{30}T^4,
\end{equation}
 with $g_*$ being the total number of relativistic degrees of freedom in the thermal plasma.
The parameter $\alpha$ is defined by
\begin{equation}
\alpha \equiv \frac{\epsilon}{\rho_\mathrm{rad}} .
\end{equation}

The bubble nucleation rate per unit volume at a finite temperature is given by
\begin{equation}
\Gamma(T) =  \Gamma_0 e^{-S(T)} \simeq \Gamma_0 e^{-S^3_E(T)/T} .
\end{equation}
Here, $\Gamma_0$ is a coefficient of the order of the transition energy scale,
 $S$ is the action in the four-dimensional Minkowski space, 
 and $S^3_E$ is the three-dimensional Euclidean action~\cite{Turner:1992tz}.
The transition timescale is characterized by a dimensionless parameter
\begin{align}
\frac{\beta}{H_{\star}} \simeq \left. T\frac{d S}{d T}\right|_{T_{\star}} = \left. T\frac{d (S^3_E/T)}{d T} \right|_{T_{\star}}  ,
\end{align}
 with
\begin{equation}
\beta \equiv -\left.\frac{d S}{d t}\right|_{t_\star}. 
\end{equation}

\subsection{GW spectrum}

Here, we briefly note formulas of generated GW by each of three sources:
 bubble collisions, turbulence, and sound waves after bubble collisions.
The final spectrum is expressed, by taking the sum of all three, as
\begin{equation}
\Omega_{GW}(f) = \Omega_{GW}^\mathrm{coll}(f) +  \Omega_{GW}^\mathrm{sw}(f) +  \Omega_{GW}^\mathrm{turb}(f),
\end{equation}
 in terms of the density parameter.
For information, we find that the bubble collision contribution is negligible, 
 the sound wave is the dominant source, and turbulence gives a high frequency tail in the spectrum,
 as GWs generated by a first-order phase transition in many other models.

\subsubsection{Bubble collisions}

The peak frequency and the peak amplitude of GWs generated by bubble collisions are,
 under the envelope approximation\footnote{For a recent development beyond the envelope approximation,
 see Ref.~\cite{Jinno:2016vai}.} 
 and for $\beta/H_{\star} \gg 1$~\cite{Kosowsky:1992vn}, expressed by
\begin{align}
f_\mathrm{peak} &\simeq 17 \left(\frac{f_{\star}}{\beta}\right) \left(\frac{\beta}{H_{\star}}\right)
 \left(\frac{T_{\star}}{10^8 \, \mathrm{GeV}}\right)\left(\frac{g_*}{100}\right)^{1/6} \mathrm{Hz} , \\
h^2 \Omega_{GW}^\mathrm{coll}(f_\mathrm{peak} ) &\simeq 1.7 \times 10^{-5} \kappa^2\Delta \left(\frac{\beta}{H_{\star}}\right)^{-2}
 \left(\frac{\alpha}{1+\alpha}\right)^2 \left(\frac{g_*}{100}\right)^{-1/3} ,
\end{align}
with 
\begin{align}
\Delta &= \frac{0.11 v_b^3}{0.42+v_b^2},\\
\frac{f_{\star}}{\beta} &= \frac{0.62}{1.8-0.1 v_b+v_b^2},
\end{align}
 where $v_b$ is the bubble wall velocity.
Here, the efficiency factor ($\kappa$) is given by~\cite{Kamionkowski:1993fg}
\begin{equation}
\kappa = \frac{1}{1+ A \alpha}\left( A \alpha +\frac{4}{27}\sqrt{\frac{3\alpha}{2}} \right),
\end{equation}
 with $A = 0.715 $.
The full GW spectrum can be well fitted by~\cite{Huber:2008hg}
\begin{equation}
\Omega_{GW}^\mathrm{coll}(f)  = \Omega_{GW}^\mathrm{coll}(f_\mathrm{peak})
 \frac{(a+b)f_\mathrm{peak}^b f^a}{ b f_\mathrm{peak}^{a+b} + a f^{a+b}},
\end{equation}
 with numerical factors $a \in [2.66, 2.82]$ and $b \in [0.90, 1.19]$.
We adopt the values of $(a, b, v_b)=(2.7, 1.0, 0.6)$ in our analysis.

\subsubsection{Sound waves}
 
The peak frequency and the peak amplitude of GWs generated by sound waves are expressed
 by~\cite{Hindmarsh:2013xza,Hindmarsh:2015qta} 
\begin{align}
f_\mathrm{peak} &\simeq 19 \frac{1}{v_b} \left(\frac{\beta}{H_{\star}}\right)
 \left(\frac{T_{\star}}{10^8 \, \mathrm{GeV}}\right)\left(\frac{g_*}{100}\right)^{1/6} \mathrm{Hz} ,
\label{Eq:Peak_Sound} \\
h^2\Omega_{GW}^\mathrm{sw}(f_\mathrm{peak} ) &\simeq 2.7 \times 10^{-6} \kappa_v^2 v_b \left(\frac{\beta}{H_{\star}}\right)^{-1}
 \left(\frac{\alpha}{1+\alpha}\right)^2 \left(\frac{g_*}{100}\right)^{-1/3}.
\label{Eq:Amplitude_Sound}
\end{align}
The efficiency factor ($\kappa_v$) is given by~\cite{Espinosa:2010hh}
\begin{equation}
\kappa_v \simeq \left\{
\begin{array}{lll}
  v_b^{6/5} \frac{6.9 \alpha}{1.36-0.037\sqrt{\alpha}+\alpha }  \quad & \textrm{for} & v_b \ll c_s \\
  \frac{\alpha}{0.73 + 0.083\sqrt{\alpha}+\alpha }       & \textrm{for} & v_b \simeq 1 \\
\end{array}
\right. ,
\end{equation}
 where $c_s$ denotes the sonic speed.
The spectrum shape is well fitted by\footnote{
We employ this formula in our analysis.
Recently, it has been claimed~\cite{Ellis:2018mja,Ellis:2019oqb}
 that the sound wave period seems to be shorter than 
 what has been expected in literature, and the resultant GW amplitudes  
 can be suppressed differently than what we have obtained in this paper. 
We leave the clarification of this issue for future study. }~\cite{Caprini:2015zlo}
\begin{equation}
\Omega_{GW}^\mathrm{sw}(f)  = \Omega_{GW}^\mathrm{sw}(f_\mathrm{peak})
\left(\frac{f}{f_{\mathrm{peak}}} \right)^3 
 \left( \frac{7}{ 4+ 3 \left(\frac{f}{f_\mathrm{peak}}\right)^2 } \right)^{7/2} .
\end{equation}

\subsubsection{Turbulence}

The peak frequency and amplitude of GWs generated by turbulence are expressed by~\cite{Kamionkowski:1993fg}
\begin{align}
f_\mathrm{peak} &\simeq 27 \frac{1}{v_b} \left(\frac{\beta}{H_{\star}}\right)
 \left(\frac{T_{\star}}{10^8 \, \mathrm{GeV}}\right)\left(\frac{g_*}{100}\right)^{1/6} \mathrm{Hz} , \\
h^2\Omega_{GW}^\mathrm{turb}(f_\mathrm{peak} ) &\simeq  3.4 \times 10^{-4} v_b \left(\frac{\beta}{H_{\star}}\right)^{-1}
 \left(\frac{\kappa_\mathrm{turb} \alpha}{1+\alpha}\right)^{3/2}
 \left(\frac{g_*}{100}\right)^{-1/3} .
\end{align}
In our analysis, we follow Ref.~\cite{Caprini:2015zlo} and
 conservatively set the efficiency factor for turbulence to be 
 $\kappa_\mathrm{turb} \simeq 0.05 \kappa_v$.
The spectrum shape is well fitted  by~\cite{Caprini:2009yp,Binetruy:2012ze,Caprini:2015zlo}
\begin{equation}
\Omega_{GW}^\mathrm{turb}(f)  = \Omega_{GW}^\mathrm{turb}(f_\mathrm{peak})
\frac{\left(\frac{f}{f_{\mathrm{peak}}} \right)^3}{
 (1 + \frac{f}{f_\mathrm{peak}} )^{11/3}(1+\frac{8\pi f}{h_{\star}})  }  ,
\end{equation}
with
\begin{equation}
h_{\star} = 17 \left(\frac{T_{\star}}{10^8 \mathrm{GeV}}\right)\left(\frac{g_*}{100}\right)^{1/6} \mathrm{Hz} .
\end{equation}
%

\section{GWs generated by the minimal $B-L$ phase transition}
\label{sec:gw}

\subsection{The minimal $B-L$ model}

\begin{table}[t]
\begin{center}
\begin{tabular}{|c|ccc|c|}
\hline
      &  SU(3)$_c$  & SU(2)$_L$ & U(1)$_Y$ & U(1)$_{B-L}$  \\ 
\hline
$q^{i}_{L}$ & {\bf 3 }    &  {\bf 2}         & $ 1/6$       & $1/3 $   \\
$u^{i}_{R}$ & {\bf 3 }    &  {\bf 1}         & $ 2/3$       & $1/3 $   \\
$d^{i}_{R}$ & {\bf 3 }    &  {\bf 1}         & $-1/3$       & $1/3 $  \\
\hline
$\ell^{i}_{L}$ & {\bf 1 }    &  {\bf 2}         & $-1/2$       & $-1 $    \\
$e^{i}_{R}$    & {\bf 1 }    &  {\bf 1}         & $-1$         & $-1 $   \\
\hline
$H$            & {\bf 1 }    &  {\bf 2}         & $- 1/2$       & $0 $   \\  
\hline
$N^{i}_{R}$    & {\bf 1 }    &  {\bf 1}         &$0$                    & $-1$     \\
$\Phi_2$            & {\bf 1 }       &  {\bf 1}       &$ 0$                  & $ + 2 $  \\ 
\hline
\end{tabular}
\end{center}
\caption{
The particle content of the minimal $U(1)_{B-L}$ model. 
In addition to the SM particle content ($i=1,2,3$), three RH neutrinos  
  [$N_R^i$ ($i=1, 2, 3$)] and one $U(1)_{B-L}$ Higgs field $\Phi_2$ are introduced.   
}
\label{table1}
\end{table}

Our model is based on the gauge group $SU(3)_C \times SU(2)_L \times U(1)_Y \times U(1)_{B-L}$ and
 the particle content is listed in Table~\ref{table1}.
In addition to the SM model particles, we introduce three RH neutrinos ($N_R^i$ with $i$ running $1,2,3$) and
 one SM singlet $B-L$ Higgs field $\Phi_2$ where the subscript 2 stands for its $B-L$ gauge charge. 
In the presence of the three RH neutrinos, the model is free from all the gauge and mixed-gravitational anomalies. 
 
The Yukawa interactions of $N_R$s are 
\begin{align}
\mathcal{L}_{Yukawa} \supset  - \sum_{i=1}^{3} \sum_{j=1}^{3} Y^{ij}_{D} \overline{\ell^i_{L}} H N_R^j 
          -\frac{1}{2} \sum_{k=1}^{3} Y_{N^k} \Phi_2 \overline{N_R^{k~C}} N_R^k 
           + {\rm H.c.} ,
\label{Lag1} 
\end{align}
 where the first term is the neutrino Dirac Yukawa coupling, and the second is the Majorana Yukawa couplings. 
Once the $U(1)_{B-L}$ Higgs field $\Phi_2$ develops a nonzero vacuum expectation value (VEV),
 the $U(1)_{B-L}$ gauge symmetry is broken and the Majorana mass terms
 of the RH neutrinos are generated. 
Then, the seesaw mechanism is automatically implemented in the model after the electroweak symmetry breaking.

We consider the following tree level scalar potential: 
\begin{align}
V_0(\Phi_2 )
 = - M^2_{\Phi_2} \Phi_2\Phi_2^{\dagger} +\frac{1}{2}\lambda_2 (\Phi_2\Phi_2^{\dagger} )^2.
\label{potential:tree}
\end{align}
Here, we omit the SM Higgs field ($H$) part and its interaction terms
 for not only simplicity but also little importance in the following discussion,
 since we are interested in the case that the VEVs of the $B-L$ Higgs field are much larger
 than that of the SM Higgs field.\footnote{For the case 
 of a phase transition of the SM Higgs field interacting with a new Higgs field,
 see, for example, Ref.~\cite{Jinno:2015doa}.}
At the U(1)$_{B-L}$ symmetry breaking vacuum, the $B-L$ Higgs field develops the VEV,
 and the RH neutrinos $N_R^i$ and the $B-L$ gauge boson ($Z^\prime$ boson) 
 acquire their masses, respectively, as 
\begin{align}
  m_{N_R^i}=& \frac{Y_{N^i}}{\sqrt{2}} v_2, \\ 
  m_{Z^\prime}^2 =& 4 g_{B-L}^2 v_2^2,
\end{align}
where $g_{B-L}$ is the $U(1)_{B-L}$ gauge coupling and $v_2$ is defined as $\langle\Phi_2\rangle = v_2/\sqrt{2}$. 
Then, the tree level $B-L$ Higgs boson mass is given as $m_{\Phi_2}^2=\lambda_2 v_2^2$.
Note the LEP constraint $m_{Z'}/g_{B-L} \gtrsim 6$ TeV~\cite{Carena:2004xs, Heeck:2014zfa} 
  and the constraint from the LHC Run 2 on the search for a narrow resonance
 (see, for example, Refs.~\cite{Okada:2016gsh,Okada:2016tci,Okada:2017pgr,Okada:2018ktp}) 
\begin{equation}
 m_{Z'} \gtrsim  3.9~\mathrm{TeV},
\label{constr:Zpmass}
\end{equation}
 for $g_{B-L}\simeq 0.7$. 

In the minimal $B-L$ model, one-loop quantum corrections to the scalar potential for both zero and finite temperature are
 essential for realizing the first-order phase transition.
For our numerical calculations, we have implemented our minimal $U(1)_{B-L}$ model
 into the public code \texttt{CosmoTransitions}~\cite{Wainwright:2011kj}, 
 where both zero- and finite-temperature one-loop effective
 potentials\footnote{As one might know, the use of the effective Higgs potential holds the issue of gauge dependence in the results~\cite{Chiang:2017zbz}. 
Since resolution to this issue is under development, we adopt the effective potential technique.}~\cite{Quiros:1999jp}, 
\begin{align}
 & V_{\mathrm{eff}}(\varphi, T) = V_0(\varphi)+ \Delta V_{1-\mathrm{loop}}(\varphi) + \Delta V_T(\varphi, T), 
\label{potential:eff} 
\end{align}
 with $\Phi_2 = \varphi/\sqrt{2}$,
 have been calculated in the $\overline{\textrm{MS}}$ renormalization scheme at a renormalization scale $Q^2=v_2^2$.
In the following calculations, we assume $Y_{N^i} \ll g_{B-L}$, for simplicity, 
 and neglect quantum corrections through neutrino Yukawa couplings $Y_{N^i}$. 
Thus, the effective potential (\ref{potential:eff}) is described by only three free parameters, 
  $g_{B-L}, \lambda_2$ and $M_{\Phi_2}$.
In our analysis, we use $v_2$ instead of $M_{\Phi_2}$. 

\subsection{Parameter dependence}

\begin{figure}[htbp]
\centering
\includegraphics[clip,width=12.0cm]{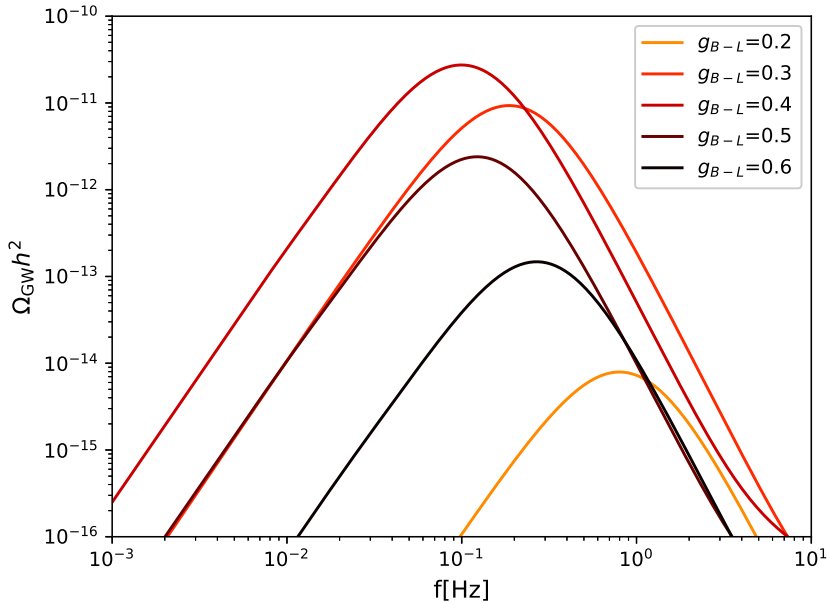}
\caption{
The predicted GW spectrum for various values of $g_{B-L}$ 
   for $v_2 = 10$ TeV and $\lambda_2 =0.002$. 
}
\label{Fig:gbl_dependence}
\end{figure}

We now show a dependence of our results on three free parameters: $g_{B-L}, \lambda_2$, and $v_2$. 
At first, we focus on the gauge coupling dependence of the resultant GW spectrum.
The GW spectrum for various values of the $B-L$ gauge coupling constant 
  for the fixed value of $v_2 = 10$ TeV and $\lambda_2 =0.002$ is shown in Fig.~\ref{Fig:gbl_dependence}.
We have found a mild dependence for the frequency but the amplitude is quite sensitive.
The largest amplitude is obtained for $0.35 \lesssim g_{B-L} \lesssim 0.4$. 

\begin{figure}[htbp]
\centering
\includegraphics[clip,width=12.0cm]{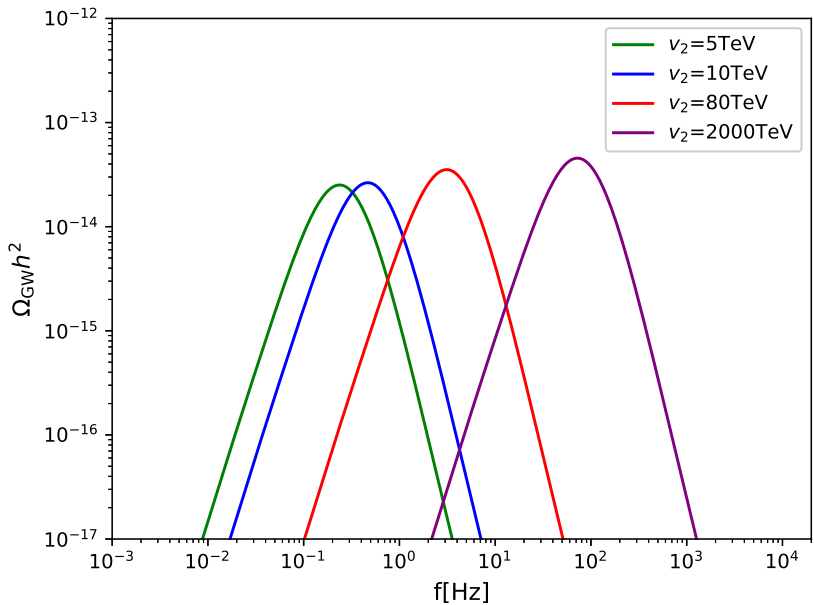}
\caption{
The predicted GW spectrum for various values of $v_2$ 
   for $g_{B-L}=0.4$ and $\lambda_2 =0.001$. 
}
\label{Fig:v2_dependence}
\end{figure}

Next, we focus on the VEV dependence of the resultant GW spectrum.
We show in Fig.~\ref{Fig:v2_dependence} the GW spectrum for various VEVs for the fixed values
 of $g_{B-L} = 0.4$ and $\lambda_2 =0.01$.
We have found a very mild dependence for the amplitude but a strong dependence of the peak frequency.
We have found an approximate relation of $f_{\mathrm{peak}} \propto v_2$.
This can easily be understood, because $v_2$ is the only dimensionful parameter and scales the system as well as $T_\star$.
Thus, we see $f_{\mathrm{peak}} \propto T_\star \propto v_2$ from Eq.~(\ref{Eq:Peak_Sound}).
%

\begin{figure}[htbp]
\centering
\includegraphics[clip,width=12.0cm]{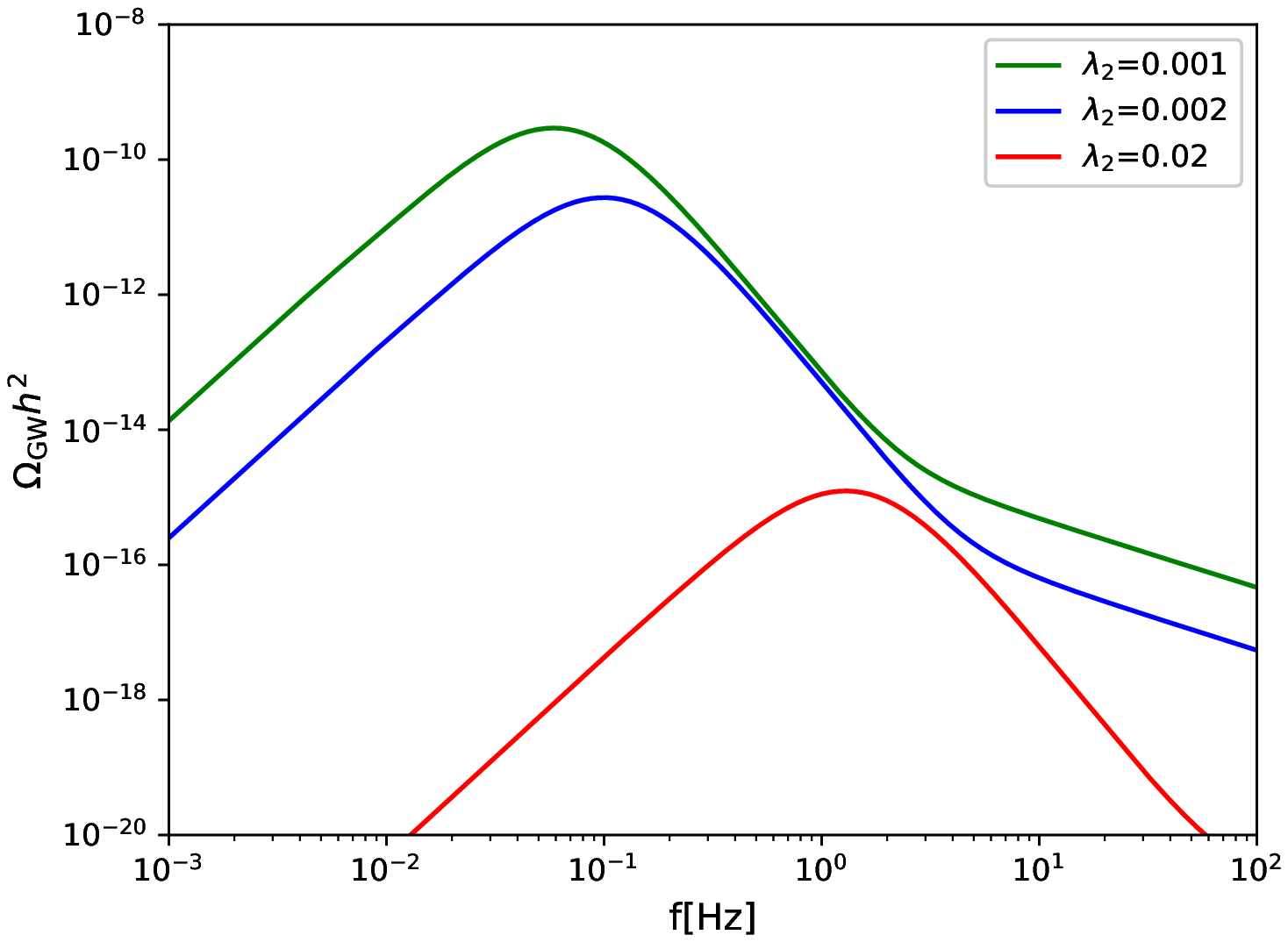}
\caption{
The predicted GW spectrum for various values of $\lambda_2$ 
   for $g_{B-L}=0.4$ and $v_2 =10$ TeV. 
}
\label{Fig:l2_dependence}
\end{figure}

At last, we focus on the $\lambda_2$ dependence of the resultant GW spectrum.
We show the GW spectrum for various $\lambda_2$ values for $g_{B-L} = 0.4$ and $v_2 =10$ TeV 
 in Fig.~\ref{Fig:l2_dependence}.
As $\lambda_2$ decreases, the peak frequency also decreases while the peak amplitude increases. 
We approximate relations such as
 $\Omega_{\mathrm{GW}}h^2(f_\mathrm{peak}) \propto \lambda_2^{-1/4}$ and $f_{\mathrm{peak}} \propto \lambda_2$.
We also find a lower bound on $\lambda_2$.
As will be listed in Table~\ref{points}, $\alpha$ is as large as $\mathcal{O}(1)$ for $\lambda_2 \sim 10^{-4}$,
 which means that the energy density of background radiation and that of the latent heat is comparable. 
For $\lambda_2 \ll 10^{-4}$, the latent heat becomes too large. 
Indeed, we find the effective equation of state parameter $w$ becomes smaller than $-1/3$
 for such a too small $\lambda_2$ for $g_{B-L} = \mathcal{O}(0.1)$.
In this case, the universe would inflate as in the old inflation model~\cite{Sato:1980yn,Guth:1980zm}.
 
%

\subsection{Predicted spectrum for benchmark points}

\begin{figure}[htbp]
\centering
\includegraphics[clip,width=12.0cm]{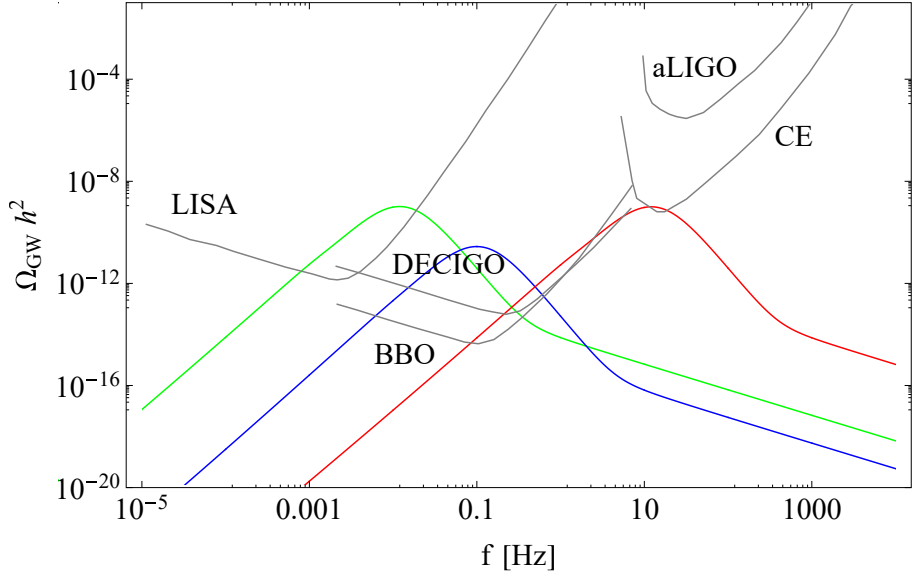}
\caption{
The predicted GW spectra for the benchmark points are shown.
Green, blue and red curves from left to right correspond to points A, B, and C, respectively.
The future experimental sensitivity curves of LISA~\cite{Sathyaprakash:2009xs}, DECIGO and BBO~\cite{Yagi:2011wg},
 aLIGO~\cite{TheLIGOScientific:2014jea}, and Cosmic Explore (CE)~\cite{Evans:2016mbw} are also shown in black. 
}
\label{Fig:GWspectrum}
\end{figure}

We list our results for three benchmark points in Table~\ref{points}.
\begin{table}
\begin{center}
  \begin{tabular}{|c|ccc|ccc|c|}
   \hline
 Point & $g_\mathrm{B-L}$ &  $v_2$ & $\lambda_2$ &  $\alpha$ & $\beta/H_{\star}$ & $T_\star$ & $\varphi_\mathrm{C}/T_\mathrm{C}$   \\\hline 
A & $0.44$ &  $4$ TeV & $1.5 \times 10^{-4}$ & $0.97$ & $115.2$ & $0.327$ TeV & $3.46$ \\\hline
B & $0.40$ &  $12$  TeV & $2.0 \times 10^{-4}$ & $0.32$ & $304.5$ & $1.032$ TeV & $3.64$ \\\hline
C & $0.46$ &  $3800$ TeV & $4.0 \times 10^{-4}$ & $0.96 $ & $115.4 $ & $328.7$ TeV & $3.30$ \\\hline
   \end{tabular}
\end{center}
  \caption{Input and output parameters for several benchmark points are listed.}
\label{points}
\end{table}
Here, the last quantity, $\varphi_c/T_\mathrm{C}$, is a typical measure of the strength of
 the first-order phase transition
with the critical temperature $T_\mathrm{C}$ being the temperature when two minima are degenerate in the effective potential and $\varphi_\mathrm{C}$ being the value of the nontrivial minimum.
In Fig.~\ref{Fig:GWspectrum}, we show predicted GW spectra 
 for our benchmark points along with expected sensitivities of various future interferometer experiments. 
Green, blue, and red curves from left to right correspond to points A, B, and C, respectively.
Black solid curves denote the expected sensitivities of each indicated experiments:
 LISA~\cite{Sathyaprakash:2009xs}, DECIGO and BBO~\cite{Yagi:2011wg},
 aLIGO~\cite{TheLIGOScientific:2014jea}, and Cosmic Explore (CE)~\cite{Evans:2016mbw}.
Curves are drawn by \texttt{gwplotter}~\cite{Moore:2014lga}.
The sensitivities of DECIGO and BBO reach the results of the benchmark points A and B.

\section{Summary}

In this paper, we have calculated the spectrum of stochastic GW radiation generated
 by the cosmological phase transition of the minimal $U(1)_{B-L}$ model.
We have found that a first-order phase transition strong enough to generate GWs with a detectable amplitude
 can be realized in the minimal $U(1)_{B-L}$ model with a single $B-L$ Higgs field,   
  while an additional Higgs field has been thought to be necessary for such a strong first-order phase transition
  through previous studies. 
The Higgs potential of the minimal gauged $U(1)_{B-L}$ model is quite simple, 
  and only three parameters are involved in our analysis. 
We clarify a dependence of the resultant GW spectrum on the three parameters:
 the peak amplitude is sensitive to the gauge coupling constant and the self-coupling constant,
 while the peak frequency is roughly proportional to the VEV of the $B-L$ Higgs field and the self-coupling constant.
The $B-L$ phase transition at an energy scale far beyond the LHC reach can be observed through GWs in the future.
We have also found, for a sensible value of the gauge coupling constant,
 the existence of a lower bound on the Higgs self-coupling constant $\lambda_2 \gtrsim 10^{-4}$
 in order not to realize an unwanted second inflation. 
We stress that, although our analysis has been done based on the $U(1)_{B-L}$ model,
 our results in this paper are general and applicable for any $U(1)$ gauge theory with a minimal Higgs sector, 
 as long as Yukawa coupling effects on the effective Higgs potential are negligible.


\section*{Acknowledgments}
We are grateful to Motoi Tachibana for a valuable conversation.
This work is supported in part by the U.S. DOE Grant No.~DE-SC0012447 (N.O.) and
 KAKENHI Grants No.~19K03860 and No.~19H05091 (O.S.).
%





\end{document}